\documentclass[a4paper]{llncs}
\usepackage{fullpage}
\usepackage{cite}
\usepackage{lineno}
\usepackage{amsmath}
\usepackage{amssymb}

\usepackage[compatibility=false]{caption}
\usepackage{subcaption}
\usepackage{graphicx}

\begin{document}

\title{Parking Sensing and Information System: Sensors, Deployment, and Evaluation}
\titlerunning{Parking Sensing System}  
%
\author{Xiao Chen\inst{1} \and Zhen (Sean) Qian \inst{2} \and
Ram Rajagopal \inst{1} \and Todd Stiers \inst{3} \and Christopher Flores \inst{3} \and Robert Kavaler\inst{3} \and
Floyd Williams III \inst{3} }
\authorrunning{Xiao Chen et al.} 
%
%
\institute{
Stanford University,
Stanford, CA 94305
\\
\and
Carnegie Mellon University,
Pittsburgh, PA 15213
\and 
Sensys Networks, Inc
Berkeley, CA 94710
}

\maketitle              

\begin{abstract}
This paper describes a smart parking sensing and information system that disseminates the parking availability information for public users in a cost-effective and efficient manner. The hardware framework of the system is built on advanced wireless sensor networks and cloud service over the Internet, and the system is highly scalable. The parking information provided to the users is set in the form of occupancy rates and expected cruising time. Both are obtained from our analytical algorithm processing both historical and real-time data, and are thereafter visualized in a color theme. The entire parking system is deployed and extensively evaluated at Stanford University Parking Structure-1
\end{abstract}
\section{Introduction}
Parking supply in the urban area does not meet the demand, which leads to many societal problems including traffic congestion, air pollution, and driver frustration. In some cities, it has been shown that up to 74\% of downtown congestion can be caused by drivers cruising for available parking spots \cite{shoup2006cruising}. A detailed study in Schwabing, Germany, estimates that traffic searching for available parking spots accounts for 20 million Euros (1 EUR = 1.236 USD based on 2006 annual average exchange rate) annually \cite{Caliskan_2006}. Moreover, it has been discovered that, over one year in a small Los Angeles business district, cars cruising for parking is equivalent to 38 individual trips around the world. Such cruising for parking burns 47,000 gallons of gasoline and produces 730 tons of carbon dioxide \cite{geng2012new}. Yet prior to solving these problems, we need to identify the key issues that result in an imbalance between parking supply and demand. Further investigations \cite{caicedo2006parking, tamrazian2015my} suggest that lack of parking information and incentives for drivers results in low efficiency for parking. For instance, due to the absence of parking information, a driver may leave a parking lot without knowing that a spot may soon become available. Also commonly seen is that a vacant spot is difficult to obtain during peak hours, while a large number of spots are available in the late afternoon or early morning. Therefore, parking information provision and incentives are critical for us to guide the parking demand in a smarter way to make the best use parking supply. \\~\\
Our motivation is effectively and efficiently matching the parking demand with its supply by providing useful information as an inexpensive way to guide drivers. Thus an intelligent parking sensing and information system is needed as a holistic solution to obtain accurate parking information, historically and in real-time. Data acquisition, management, and visualization need to be designed carefully to build an agile, reliable and informative system. In particular, several issues need to be addressed. 
\textbf{Flexibility:} The system should be easily scalable with simple installation/removal and configurations.
\textbf{Reliability:} The information obtained from the system should be consistent with the real world, both historically and in real-time. 
\textbf{Simplicity:} The communication between the system and users should be efficient and simple without losing information accuracy. 
\textbf{Sustainability:} The system should have a low deployment and maintenance cost. The system should support smart parking decisions on system planning and operations that are environmentally friendly. \\~\\
Our solution for building a parking sensing and information system, named Stanford Smart Parking (SSP), is based on Wireless Sensor Networks (WSN), cloud service and Mobile Technology. It collects spot-by-spot parking data through WSN, then transmits the compressed data to the cloud server through the cellular network, and finally feeds parking information to users through a mobile interface, which relies on our data analytics framework. The system along with the data analytics framework contributes to the parking research in the following aspects:
\begin{enumerate}
\item It is easy to be deployed and maintained with low cost.
\item	It provides robust and useful information to users.
\item	It utilizes sampled measurements (i.e. top floor data) to infer the information about the whole parking structure.
\item	It enables demand-responsive pricing strategy to manage parking demand dynamically.
\end{enumerate} 
The remainder of the paper is organized as follows: The related work and review of existing literature is presented in Section 2. Then the architecture of the system is described in Section 3. Section 4 introduces system deployment and Section 5 will evaluate the system performance. Section 6 concludes the paper.
\section{Related Works} \label{S2:related_work}
Wireless Sensor Networks with cloud service have been implemented for parking management. Tang et al. \cite{tang2006intelligent} developed a parking system based on Crossbow Mote products and Crossbow XMesh network architecture. Srikanth et al. \cite{srikanth2009design} implemented a Smart Parking system using MTS310 and MicaZ motes for sensing and detection. Both of them use WSN as the data acquisition technology. But neither of them provide web interface for users to get parking information conveniently. Yan et al. \cite{yan2011smartparking} proposed a parking system with wireless transceivers, parking belts, infrared devices and control computers. Although it partially addresses security and privacy issues, the deployment cost is high and the system does not seem scalable. Additionally, the user interface is not straightforward at first glance. Souissi et al. \cite{souissi2011parking} also developed a similar system based on CC2430 (a ZigBee \cite{baronti2007wireless} application) and another base station together with corresponding web interfaces. However, the system scale is small with only 5 sensor nodes. Lee et al. \cite{lee2008intelligent} implemented a counting sensor at a parking infrastructure’s entrance and exits as the data collection component. Although it reduced the installation/maintenance cost substantially by reducing the number of sensor nodes, the information is not accurate and prompt enough for users because they may not be aware that a parking spot has become available until after a car has left the infrastructure. It also had a significant drawback due to the counting bias accumulating over time that requires periodic reset of the occupancy. Panayappan et al. \cite{panayappan2007vanet} proposed a parking system in VANET to locate the available parking lots and open spots. This system used roadside units to relay parking messages and GPS to locate vehicle position. The parking data is aggregated at the city level and its practical implementation remains uncertain. More importantly, none of the studies incorporate any data analytics framework to provide information to users that is straightforward, easy to understand and useful. \\~\\
In the SSP, the simple but reliable WSN technologies, such as the Microradar Sensors, the Repeaters, and the Access Point \cite{haoui2008wireless}, are used as data acquisition components. Moreover, data management and data visualization components are established based on a cloud server resulting in flexible and scalable configurations and integration.
\section{System Architecture} 
\begin{figure}[h!]
\centering
\includegraphics[width=0.8\textwidth]{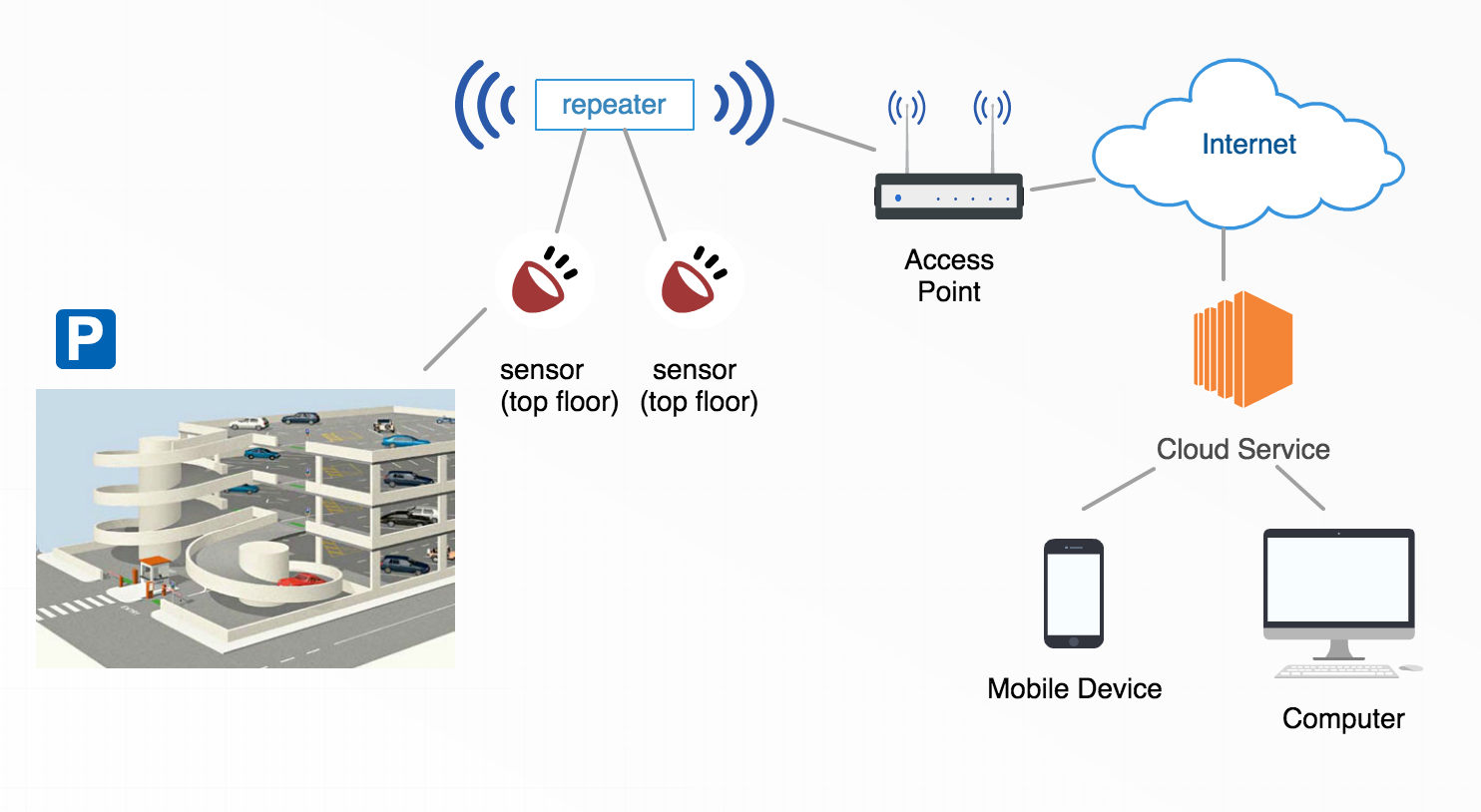}	
\caption{System Architecture}
\label{fig:sys_arch}
\end{figure}

Our system, referred to the previous system configuration in \cite{haoui2008wireless}, is yet designed to be highly scalable for each component. The flexibility is obtained by carefully designing the structure of the Wireless Sensor Network and Cloud Service platform. The reliability is achieved by having proper installation of sensors and configuration of the communication from sensor nodes to a cloud server. The simplicity is based on a high performance web platform that creates informative and simple data visualization. Sustainability is achieved from various aspects including the easy deployment, the low-power consumed mode, and potentials to reduce parking and traffic congestion. Figure \ref{fig:sys_arch} depicts the high- level architecture of the whole SSP system. Inferring the whole parking structure occupancy based on top floor measurements is reasonable when a parking structure is typically filled from the bottom to the top. For the top floor, traditional sensors do not work well because they or their wires need to be mounted on ceilings. Therefore the wireless sensors are preferred in this setting. One of our major innovations is that we utilize the behavioral principle to measure parking and therefore can substantially reduce the system cost. To our best knowledge, this is the first work that introduces the concept of inferring the overall occupancy in a garage based off spot-by-spot parking measurements on the top floor.\\~\\

\subsection{Data Acquisition} 
\subsubsection{Wireless Sensor Node} 
A wireless node is a battery-powered sensor system with microprocessor, wireless transceiver, and memory \cite{liao2014snowfort}. With its specific configuration, the data is collected, compressed and transmitted to a base station (i.e. access point) or repeater. Many Wireless Sensor Network systems use magnetic sensors, which can fail when a car parks for hours. We completed field tests and chose Microradar sensor that offers the best detection accuracy. Specifically, our system uses VSN240-M Microradar \cite{kavaler2012micro,Sensor240M} as the individual sensor node to detect weather a vehicle is parked or not. It has a programmable detection range between 4 feet and 10 feet. The width of a detection zone is approximately 90 degrees. It carries a battery with a 10-year lifetime and can be installed in less than 10 minutes \cite{haoui2008wireless}. Each sensor node operates one of 5 modes: $\mu$A (Detect), $\mu$ B (IF Display), $\mu$ C (Debug), $\mu$D(Parking), $\mu$ E(Idle). In our implementation, mode $\mu$D is set to detect the parking event. Other applications include bicycle detection or vehicle identification \cite{Sensor240M}. 

\subsubsection{Repeater} 
The repeater is designed to amplify and transmit a predetermined received signal. A series of repeaters is capable of relaying information in stringent wireless-propagation environments by forming an ad-hoc network \cite{oh2014wave}. In our system, RP240-BH Repeater \cite{Rp240BH} is chosen to provide two-way relay between out-of-range sensors and the access point without any wires or cables. The battery lifetime is about 2 years with only tens of dollars for its material and replacement cost.
\subsubsection{Access Point}
The Access Point (AP) is an intelligent device operating under the Linux operating system that maintains two-way wireless links to vehicle detection sensors and repeaters. It is DC-powered and can support standard POE (power over Ethernet) powering or battery. It performs as the base station of the local wireless sensor network. In our system, AP240, together with an Access Point Controller Card (APCC), plays a base station role. It extracts information from each sensor node every 30 seconds and transfers the compressed data to the remote server through integrated General Packet Radio Service (GPRS) cellular data modem via Global System for Mobile (GSM) Communications.
\subsection{Data Management} 
\subsubsection{Communication} 
The radio that communicates in local sensor networks is based on IEEE 802.15.4 PHY standard. It operates on one of 16 channels of 5MHz in the 2.4-2.48 GHz band \cite{haoui2008wireless}. The whole internal connection uses the Sensys Nanopower Protocol, which is based on the time division multiple access protocol. The external communication from the access point to the remote server is based on transmission control protocol–internet protocol, which supports protocols including Telnet, file transfer protocol, hypertext transfer protocol (HTTP), point-to-point protocol, and point-to-point tunneling protocol. The AP, acting as a gateway, also supports TCP streaming to applications, providing transparent access to sensor data. It can supply synchronization to applications by asking for SNP packets.
\subsubsection{Cloud Service} 
Cloud Service is an integrated server solution to provide connectivity to AP in the air, including data archiving, data processing, data storage, and performance reports. In our configuration, the cloud server mainly performs two tasks: data archiving and data processing. The server receives the data from the AP and handles requests sent from users. It communicates to the AP through HTTP protocol and supports user requests in a format of RESTful (representational state transfer) web service. The data storage module is deployed in two mechanisms-a database and static files. The database records the parking status changing point of each sensor node, which creates an efficient way to generate real-time status for the whole parking lot. Currently MySQL is the database that we chose due to its open source license, simplicity of manipulation, and stable functionality. The static files are currently stored in an extended drive by running a perl and shell mixed program periodically. Each date has an individual file. Such a file storage mechanism is highly scalable in the Hadoop file system and will empower our analytics framework with high performance in the future. The data processing module is achieved in PHP scripts. The typical data format such as JSON or XML is generated to support public API. A data verification algorithm is also embedded in the data processing module to verify the integrity of the data.
\subsection{Data Visualization}
An intuitive way of visualizing parking information will enable better parking and travel experience for users. We designed a color scheme to represent occupancy rates and corresponding cruising time that provides users a straightforward understanding of how the parking saturation looks like in the real time.
We used a web interface as the public panel to support both desktop and mobile inter- actions. The SSP system can be accessed any time from various locations with different devices. It allows users to view the real time information as well as future predicted information conveniently. We utilized JQuery, CSS3, and HTML5 as the front-end framework because it is lightweight in terms of configuration and easy to extend. A cruising time estimation is computed based on our real-time algorithm so that users are able to access useful information dynamically.
Moreover, the advantage of such an interface lies in creating a better travel experience for travelers in the future. By incorporating mobile GPS functionality and knowing where precisely the travelers are, they will be able to get more accurate travel time prediction, including parking search time. Furthermore, travelers can plan their trips ahead of time to reduce their travel time and cruising time.
\section{DEPLOYMENT}
Our parking system has been deployed in the Stanford University Parking Structure 1(PS-1). A total of 107 wireless sensors are implemented on the top floor of the PS-1. 103 of them are parking detection sensors and the remaining 4 of them are counting sensors.
\subsection{Device Installations and Calibrations}
\begin{figure}[h!]
\centering
	\begin{subfigure}[t]{0.45\textwidth}
		\includegraphics[width=\textwidth]{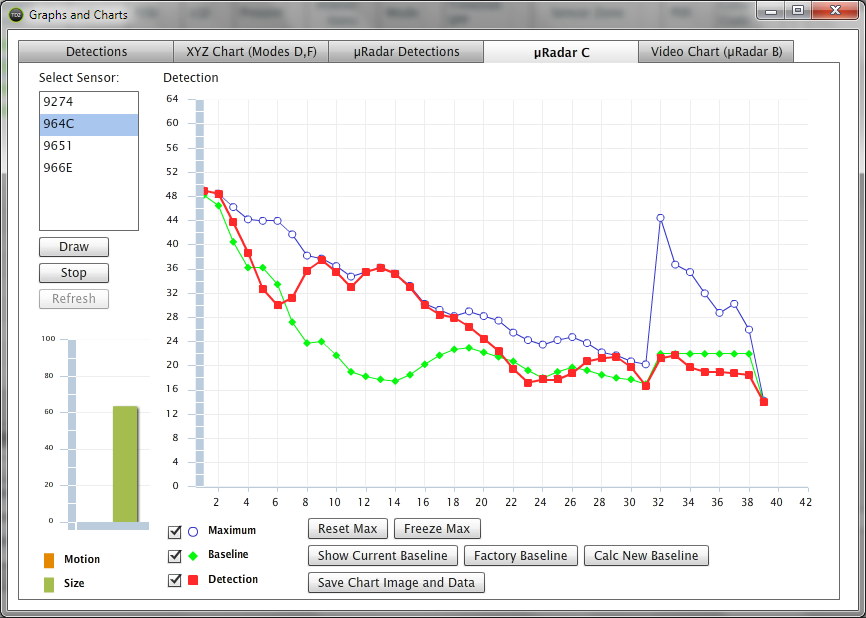}
		\caption{a car parked}
	\end{subfigure}
	\begin{subfigure}[t]{0.45\textwidth}
		\includegraphics[width=\textwidth]{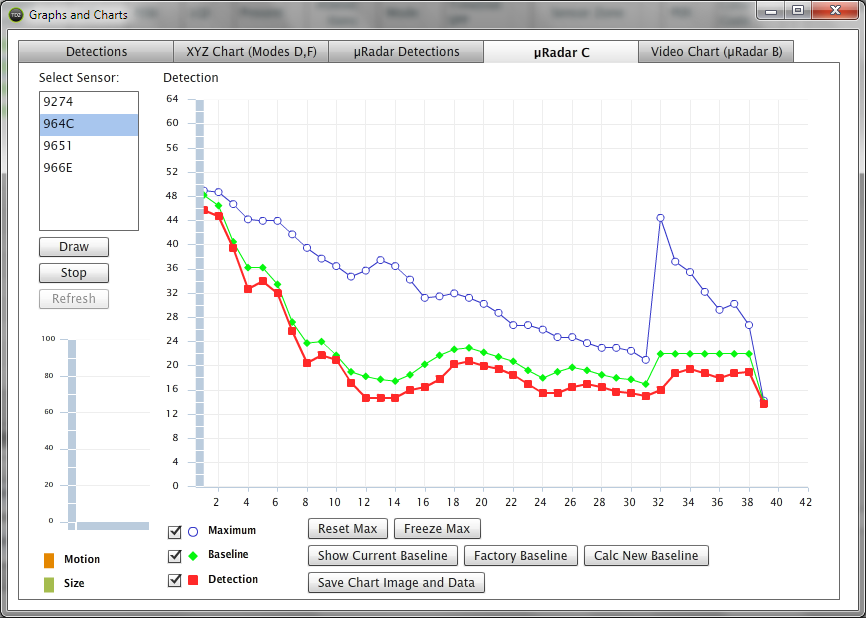}
		\caption{a car left}
	\end{subfigure}
\caption{TrafficDOT user interface, Mode C is chosen to calibrate sensors, the red line is the detection, blue line is maximum curve, green line is the baseline. From the above figure, we can see that in 4inch distance is the offset of detection range, which means the effective bins are starting from 5th point. As long as the difference between the red line and green line goes over certain threshold, namely 6dB in this case, it assumes having an object.}
\label{fig:trafficdot_cal}		
\end{figure}

A sensor is intensively tested in multiple locations to identify the best spot to detect the presence of a vehicle before all wireless sensors are deployed. In our PS-1 experiment, it was determined that, for each parking spot, the best placement for a sensor node is at the center of a parking spot, up on the curb off of the pavement. This can be done with little effort, without digging a hole in the concrete. It can be removed easily as well without any construction damage. Each sensor is buried in a plastic box covered by water-proof epoxy and glued on the curb of the pavement. It can also be buried into the ground if we want to protect it from being stolen. There are seven repeaters in our parking system at PS-1. Being mounted on top of the light poles, all of the repeaters are roughly around 14 feet high off the ground and transmit data from/to all 107 sensors. They are tilted at the proper angle to have a wide transmitting range. Each repeater communicates with 15-17 devices on an individual channel. An Access Point is installed on top of the elevator shaft in PS-1. It is about 14-15 feet high and is powered directly by an Ethernet cable. A local extended RJ45 port is also left outside for the ease of configuration of the AP if the wireless connection is not available. Inside the AP, a T-Mobile SIM (Subscriber Identification Module) card is placed to relay data from the AP to our cloud server through internet. All detection sensors are setup in parking Mode (Mode $\mu$D). In this mode, we collect data from 103 sensors which record this data every 30 seconds. Each record contains 20 bits of information including SensorID, timestamp, state, and watchdog (an indicator representing whether the state is stable or not). It is encrypted by eventproxy (a binary format) to save the file size as well as the battery of the sensor. The best thresholds of measurement parameters for the vehicle detection are 4 inches of the effective distance and 6dB of the noise level. The TrafficDOT2 (a Sensysnetworks software) is used for hardware calibration. Figure \ref{fig:trafficdot_cal} displays the visual result during the calibration.

\subsection{ Software Setup}
\begin{figure}[!b!h!t]
\centering
	\begin{subfigure}[t]{0.30\textwidth}
		\includegraphics[width=\textwidth, height=66mm]{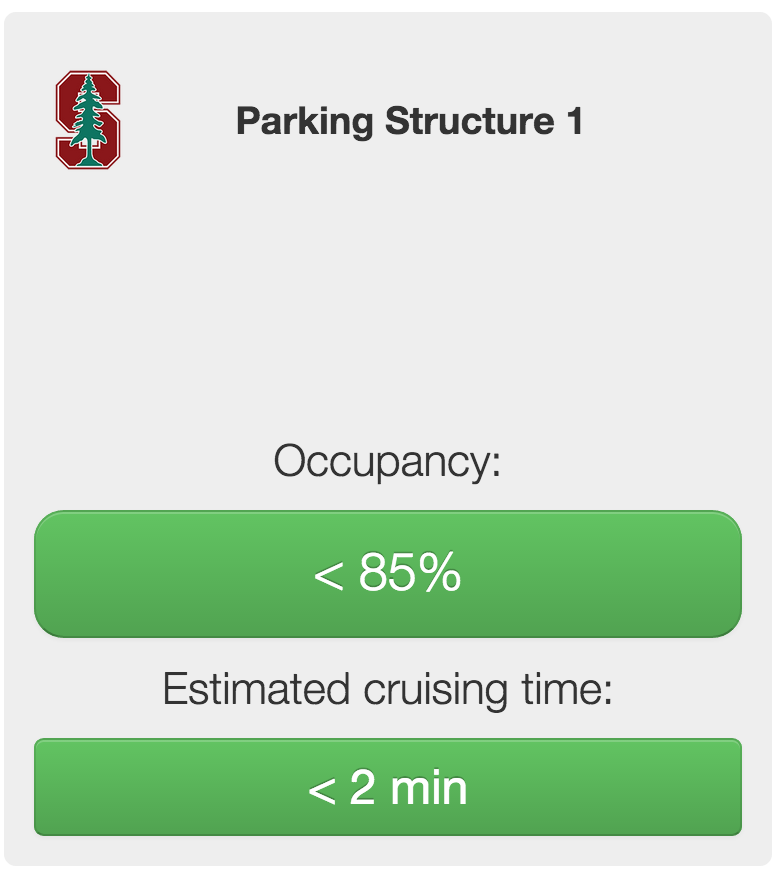}
		\caption{mobile interface case1: low occupancy}
		\label{fig:mobile_c1_g}	
	\end{subfigure}
	\begin{subfigure}[t]{0.30\textwidth}
		\includegraphics[width=\textwidth, height=66mm]{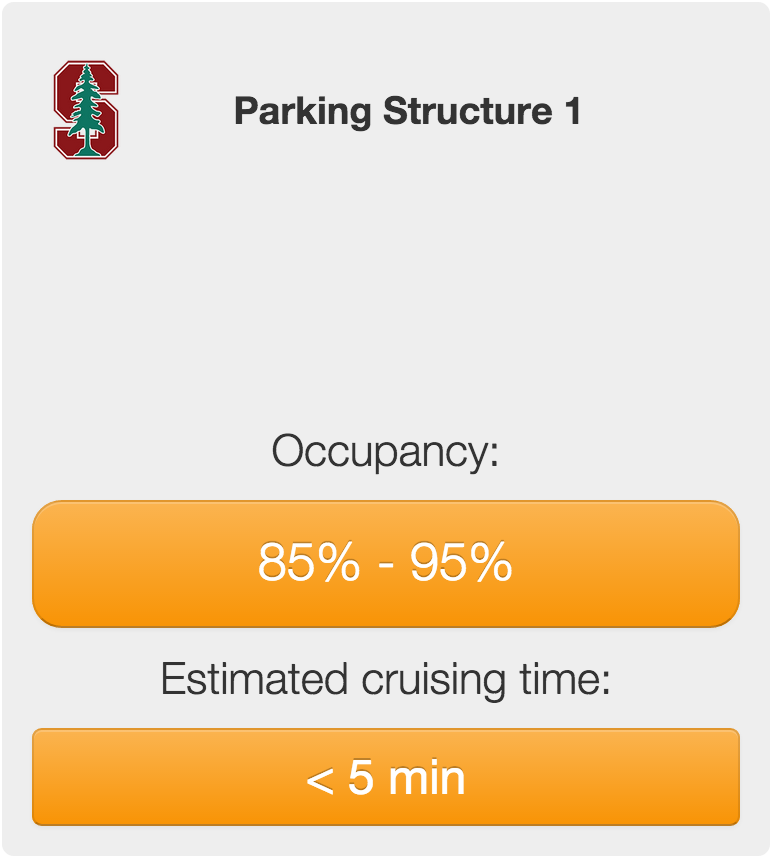}
		\caption{mobile interface case2: medium occupancy}
		\label{fig:mobile_c2_o}	
	\end{subfigure}
	\begin{subfigure}[t]{0.30\textwidth}
		\includegraphics[width=\textwidth, height=66mm]{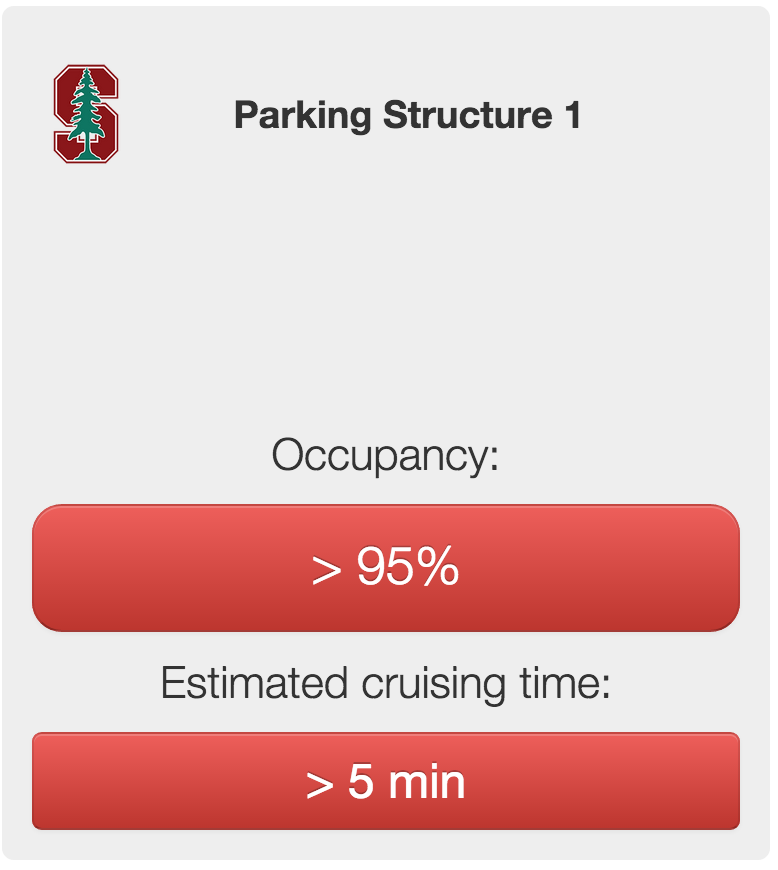}
		\caption{mobile interface case3: high occupancy}
		\label{fig:mobile_c3_r}	
	\end{subfigure}
\caption{web interface}	
\label{fig:web_interface_mobile}	
\end{figure}

The micro tier of Amazon Elastic Cloud Computing Service (also known as EC2) \cite{AWSEC2} is used as our web server. It has a 720MHz-1GHz single-core CPU and 613 MB memory. It contains 2 pieces of Amazon Elastic Block Store (EBS). One EBS is the core server that hosts the Ubuntu 12.04 operating system (Linux) and the Apache HTTP Server, which runs the various web services that supports the system \cite{fielding1997apache}. The other EBS keeps the static files of parking status on the daily basis, which facilitates our future analytics framework. The database is built on Amazon Relational Database Service (Amazon RDS) \cite{AWSRDS}. Specifically, a MySQL server, having 10 Gigabyte capacity with customized service is running in the platform. In the DB, each record contains the following content: SensorID, previous timestamp, current timestamp, original status, and updated status. Previous timestamp means the last valid time that a sensor status was changed. Similarly, current timestamp means the new valid time that this sensor status was updated. Original status and updated status are binary formatted data with 1 representing when a car is currently parked in the spot and 0 meaning no car is in that spot. Figure \cite{fig:Cloud_all_archi} shows the detailed architecture of the cloud service. Given the situation that some data may be lost or erroneous (e.g. due to communication problems or sensor malfunction in short time), some suitable algorithms from \cite{smith2003exploring} are also applied for filling the missing data or filtering the noisy data. As a result, the mobile interface is displayed in Figure \ref{fig:web_interface_mobile}. Green, Orange, and Red are 3 colors representing ease of finding available parking from easy to medium to hard. The occupancy ratio range is calculated according to the real-time sensor data, and the estimated cruising time is obtained by running our analytical algorithm in the background based on the dynamic parking information. More details about the information disclosure mechanism can be found in Section 5.
\begin{figure}[h!]
\centering
	\begin{subfigure}[b]{0.45\textwidth}
		\includegraphics[width=\textwidth]{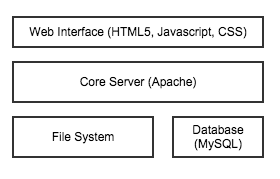}
		\caption{AWS Architecture}
		\label{fig:AWS_detail}
	\end{subfigure}	
\caption{cloud service architecture}	
\label{fig:Cloud_all_archi}	
\end{figure}

\subsection{ Capital and operational costs} 
The hardware costs a bit more then \$50k including the ground sensors, the repeaters and the base station. It costs less than existing smart parking systems used in field tests, which had only 6 sensors \cite{rodier2005transit}. The operational cost includes the deployment and maintenance cost. To deploy the whole system, it costs \$10k including installation, labor and other related costs. The maintenance cost includes the \$30 cellular data connection fee each month, and about few tens of dollars on cloud service expense each month.
\section{EVALUATION} 
Besides the reliable sensing system, the hallmark of our system is the robust data analytics framework that includes inferring/displaying parking occupancy rates and estimating parking search time accurately.
\subsection{The color scheme representing occupancy rates}
\begin{figure}[!h!b!t]
\centering
	\includegraphics[width=0.65\textwidth]{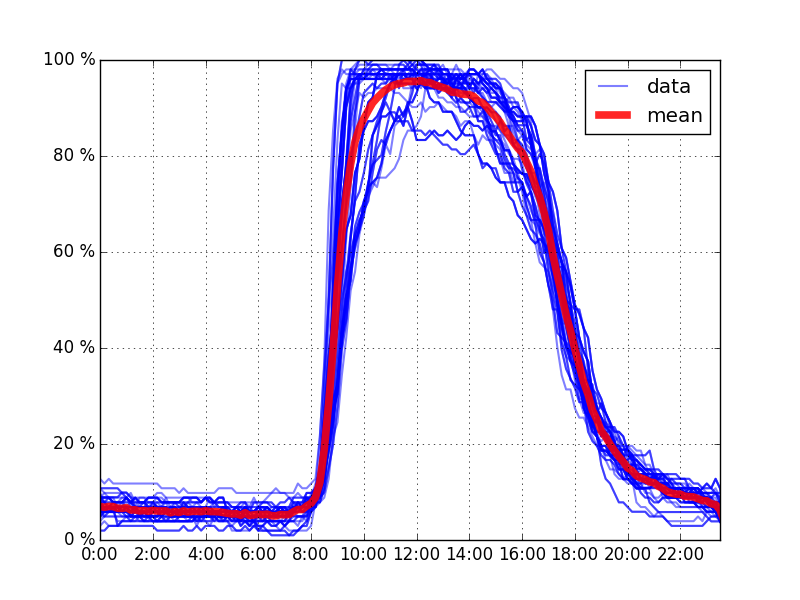}
	\caption{Weekday occupancy rate from August 01, 2013 to October 29, 2013}
	\label{fig:top_occ_ratio}
\end{figure}

\begin{figure}[!h!b!t]
\centering
\includegraphics[width=0.65\textwidth]{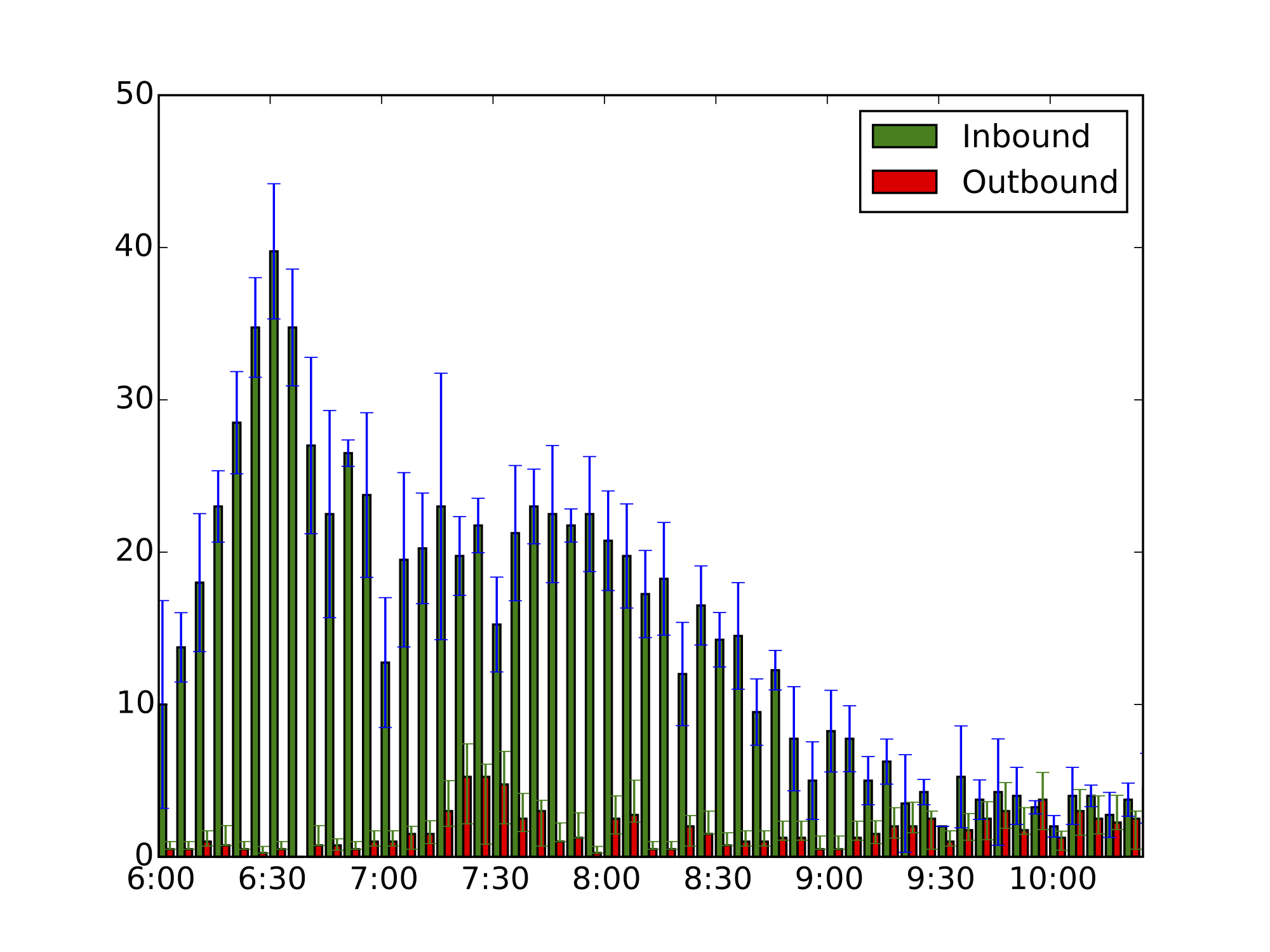}
\caption{Counts in-bound and out-bound vehicles of PS-1, from 6:00 am to 10:30am }
\label{fig:PS1_in_out_counts} 	
\end{figure}

\begin{figure}[!h!b!t]
\centering	
\includegraphics[width=0.65\textwidth]{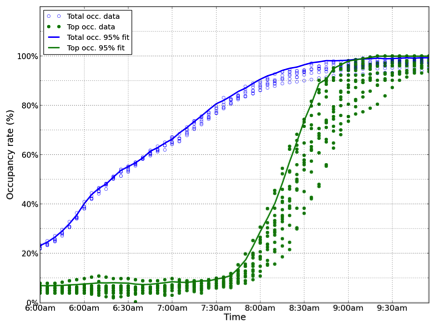}
\caption{Occupancy rate plot}
\label{fig:ps1_occ_ratio_3lines}
\end{figure}

\begin{figure}[!h!b!t]
\centering
\includegraphics[width=0.65\textwidth]{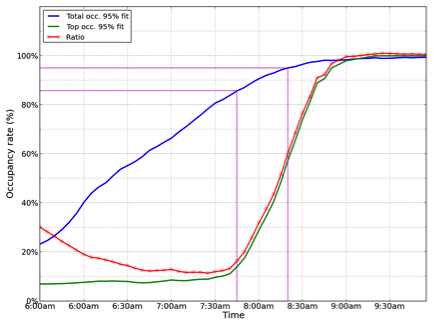}
\caption{The green line represents top occupancy rate , the blue line depicts the total occupancy rate, the ratio = $\frac{r_{t}}{r_{w}}$ is the red line } 	
\label{fig:ps1_occ_ratio_fitting}
\end{figure}

\begin{figure}[!h!b!t]
\centering
\includegraphics[width=0.65\textwidth]{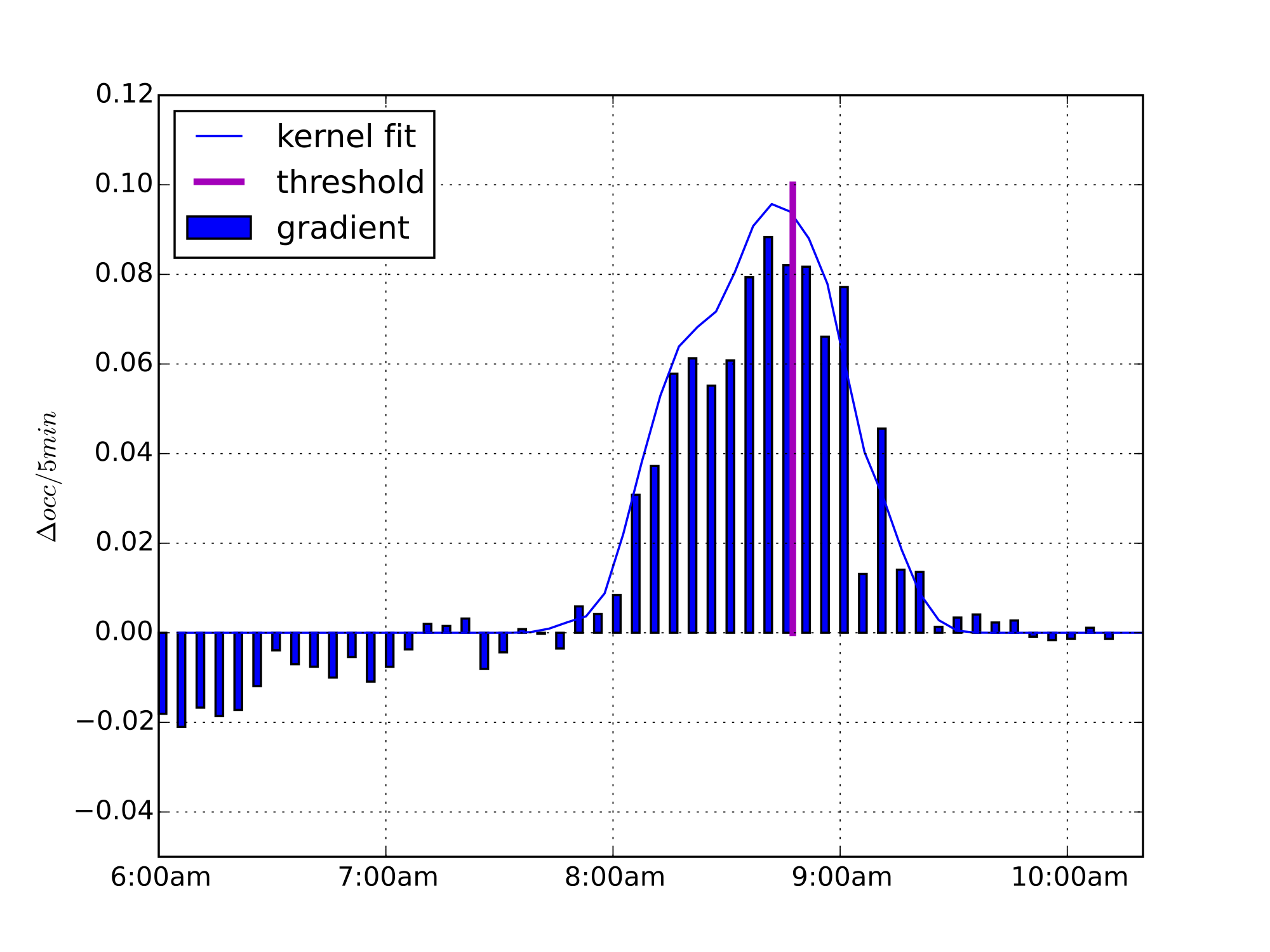}
\caption{ratio curve gradients histogram with Gaussian kernel fitting, threshold is highlighted}
\label{fig:ps1_ratio_fit_2nd_thres} 	
\end{figure}

\begin{figure}[!h!b!t]
\centering
\includegraphics[width=0.65\textwidth]{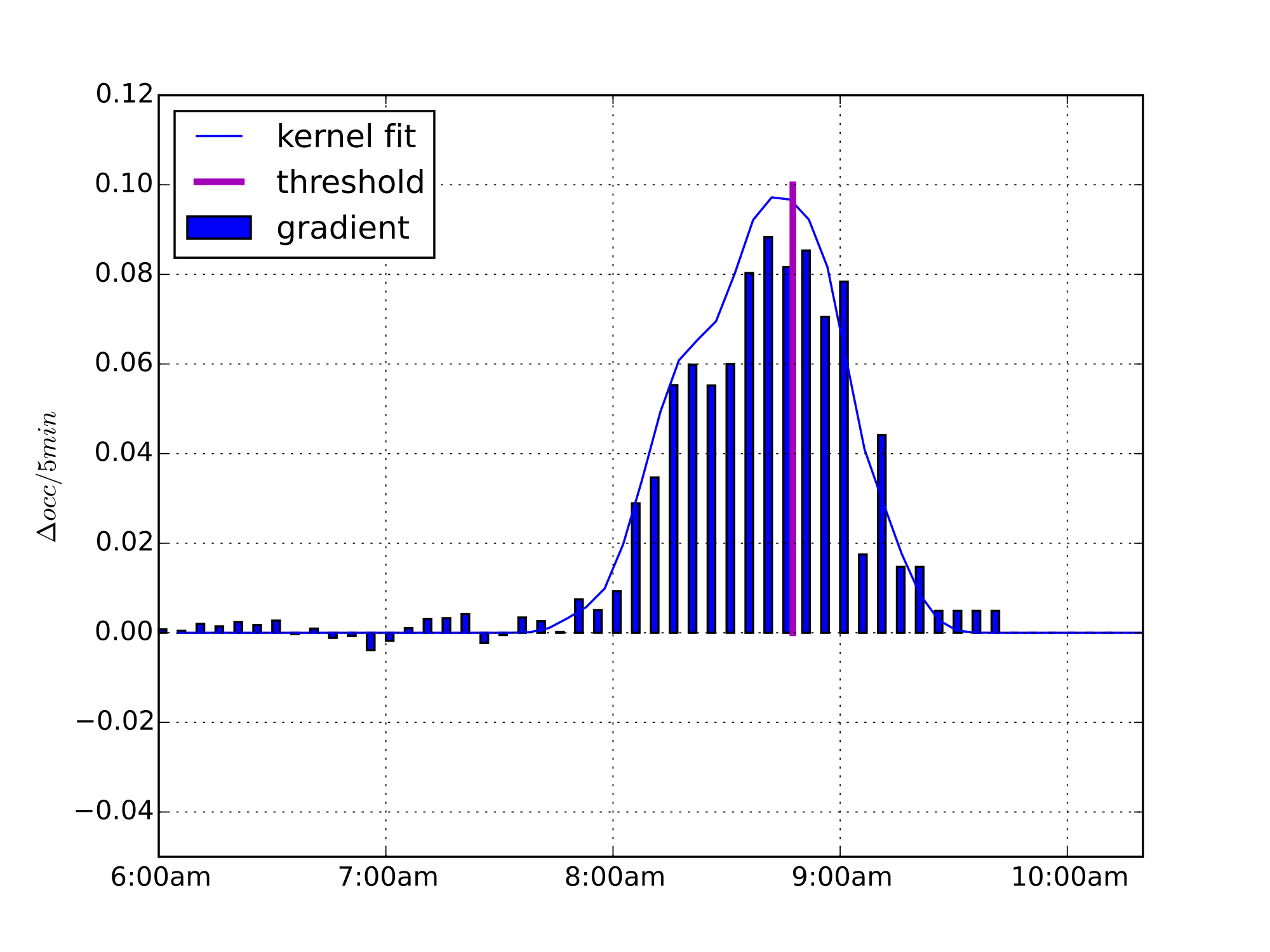}
\caption{top floor occupancy rate curve gradients histogram with Gaussian kernel fitting, threshold is highlighted}
\label{fig:ps1_Topocc_fit_2nd_thres} 	
\end{figure}

The Stanford PS-1 has in all 1080 parking spots. 928 of them are A-permit spots, which are scattered throughout the 5-floor building. The rest are A-carpool permit spots or visitor permit spots, which are all allocated on the first ground floor. The top level has 103 spots armed with 103 sensors individually; hence the top floor occupancy data is collected by our system automatically. The occupancy ratio from August 01, 2013 to October 29, 2013 is depicted in Figure \ref{fig:top_occ_ratio}. Based on the current top-floor setting, however, we would like to infer the entire PS-1 occupancy from these sampled measurements. Moreover, a green-orange-red color display scheme is introduced by SSP (Stanford Smart Parking), which enables an intuitive and prompt perception of parking congestion. To illustrate how we align the top floor occupancy with the PS1 occupancy and set up the color scheme threshold, we conducted an on-site survey spanning two weeks that counted the in-bound and out-bound vehicles in PS-1 from 6:00 am to 10:30 am. Figure \ref{fig:PS1_in_out_counts} shows the morning peak hour traffic flow through the ground floor entrance/exit for PS-1. We also pick the same time duration to show the occupancy relation between the top floor and whole PS-1. Figure \ref{fig:ps1_occ_ratio_3lines} describes the occupancy rate of the whole PS-1 and its top floor in blue and green dotted lines, respectively. A 95\% upper quantile line is fitted in the sample data due to the conservative display purpose. In other words, we chose an upper bound line (95\%) to design our color display scheme, showing that the parking structure is almost fully occupied. Figure \ref{fig:ps1_occ_ratio_fitting} shows the thresholds that display the particular color representing the associated occupancy level. The first threshold line points to the 10\% occupancy rate for the top floor, which corresponds to 85\% occupancy rate for total parking occupancy. The reason to choose 10\% occupancy on the top floor as the threshold, given that the granularity of the time series of occupancy is 5 minutes, is that the increment of ratio line (red line in Figure \ref{fig:ps1_occ_ratio_fitting}) with 10 minutes of backward span (3 consecutive time steps) goes over 50\%. The second threshold is the line that marks the 60\% occupancy rate of the top floor, which at the same time has about 95\% occupancy rate for the whole PS-1. We pick such a point as the second threshold because the ratio curve has the largest gradient value at one time step before the threshold time step point. Figure \ref{fig:ps1_ratio_fit_2nd_thres} displays the empirical result for the ratio curve (the red line the Figure \ref{fig:ps1_occ_ratio_fitting}). Furthermore we found that the ratio curve pattern is very similar to the top floor occupancy curve pattern during morning peak time, namely between 8:00 am to 10:30 am in this case. Thus we can use top-floor occupancy to present the whole PS-1 occupancy pattern during morning peak hours. Figure \ref{fig:ps1_Topocc_fit_2nd_thres} describes the threshold that we selected by picking one step ahead after the largest gradient value based on top-floor occupancy data. It reflects that a 60\% occupancy rate on the top floor is a critical point that indicates change/increment (gradient) value as the occupancy rate becomes smaller afterward. 

\subsection{False Alarm \& Mis-detection Validation}
By setting 85\% and 95\% as the thresholds of whole occupancy, we formulated the following hypothesis testing to find the critical points for the top-floor occupancy. We denote the whole parking structure occupancy rate as $r\_w$, the top-floor occupancy as $r_t$. The top-floor occupancy thresholds are defined as $\delta_1$ and $\delta_2$ to represent the critical points changing from green to orange and orange to red respectively. We have Table \ref{table:FA_MD} to depict the false alarm with respect to the missed detection.
\begin{table}[!h!b!t]
\caption{False Alarm v.s. Missed Detection}
\label{table:FA_MD}
\centering	
\begin{tabular}{|c|ccc|}
\hline
& $ 0 \leq r_w \leq 85\% $ & $ 85\% < r_w \leq 95\% $ & $ 95\% < r_w \leq 100\% $ \\
\hline
$ 0 \leq r_t \leq \delta_1 $	 & Correct & Missed Detection & Missed Detection \\
$ \delta_1 < r_t \leq \delta_2 $	 & False Alarm & Correct & Missed Detection \\
$ \delta_2 < r_t \leq 1 $	 & False Alarm & False Alarm & Correct \\
\hline
\end{tabular}
\end{table}
With those three categories of whole PS-1 occupancy rate $r_w$,  and top-floor occupancy rate $r_t$, we denote the probability of the false alarm and the missed detection as $P_{FA}$ and $P_{MD}$. To be more specific, $P_{MD\{1\}}$ and $P_{MD\{2\}}$ are the missed detection probabilities of the displayed information when the total occupancy from 0.85 to 0.95 and from 0.95 to 1 respectively. Correspondingly, $P_{MD\{1\}}$  and $P_{FA\{2\}}$ are the false alarm probabilities of the displayed information when the total occupancy is less than 0.85 and less than 0.95 respectively. To describe them in mathematical expressions, we have
\begin{align}
	P_{MD\{1\}} & = P(  \{ r_t \leq \delta_1 \} | 0.85 < r_w \leq 0.95 )  \approx \frac{ \sum_i^n 1\{ r_t^{(i)} \leq \delta_1, 0.85 < r_w^{(i)} \leq 0.95\} } {\sum_i^n 1\{ 0.85 < r_w^{(i)} \leq 0.95 \} }
\end{align}
\begin{align}		
	P_{MD\{2\}} & = P(r_t \leq \delta_2 | r_w > 0.95 ) \approx \frac{ \sum_i^n 1\{ r_t^{(i)} \leq \delta_2, r_w^{(i)} > 0.95 \} }{\sum_i^n 1\{ r_w^{(i)} > 0.95 \} }
\end{align}
\begin{align}		
	P_{FA\{1\} } & = P( r_t > \delta_1 | r_w \leq 0.85 ) \approx \frac{\sum_i^n 1\{ r_t^{(i)} \leq \delta_1, r_w^{(i)} \leq 0.85 \} }{ \sum_i^n 1\{ r_w^{(i)} \leq 0.85  \} }	
\end{align}
\begin{align}			
	P_{FA\{2\}} & = P(r_t > \delta_2 | 0.85 < r_w \leq 0.95 ) \approx \frac{\sum_i^n 1\{ r_t^{(i)} \leq \delta_2, r_w^{(i)} \leq 0.95 \} }{\sum_i^n 1 \{r_w^{(i)} \leq 0.95 \} }
\end{align}
, where $1$ is the indicator function, superscript $i$ means $i$-th recorded occupancy sample, and $n$ is the total number of occupancy samples. We pick the optimal threshold $\delta^{*} = [\delta_1^*, \delta_2^*]$ by minimizing the following cost of missed detection together with the false alarm. 
\begin{align}
	\delta^{*} = \arg \min \mathbf{C}^T \mathbf{P} \\
	s.t. \quad \mathbf{P}_{FA} \leq  \boldsymbol{\alpha}
\end{align}
where $\mathbf{P} = \begin{bmatrix}
	P_{MD\{1\} } & P_{FA\{1\}} \\
	P_{MD\{2\} } & P_{FA\{2\}}
\end{bmatrix} $,  $\mathbf{C} = \begin{bmatrix}
	C_{MD\{1\} } & C_{FA\{1\}} \\
	C_{MD\{2\} } & C_{FA\{2\}}
\end{bmatrix} $, $\mathbf{P}_{FA} = \begin{bmatrix}
	P_{FA\{1\}} \\
	P_{FA\{2\}}
\end{bmatrix} $, and $\boldsymbol{\alpha} = \begin{bmatrix}
	 \alpha_1 \\
	 \alpha_2
\end{bmatrix} $.  
$\mathbf{C}^T$ is denoted as $\mathbf{C}$ transpose.We set the cost $\mathbf{C}$ by the notion of $C_{MD\{2\} } \geq C_{MD\{1\}} \geq C_{FA\{2\} } \geq C_{FA\{1\}} $ due to the fact that the missed detection will cause more damage than the false alarm in terms of the parking information disclosure. Therefore we assign constant values to $\mathbf{C} = \begin{bmatrix}
	7 & 4 \\
	10 & 5
\end{bmatrix} $ and the false alarm tolerance level $\boldsymbol{\alpha} = \begin{bmatrix}
	0.05 \\
	0.05
\end{bmatrix}$. The resulting $\delta^* = [0.12, 0.64]^T$, which sets the top-floor occupancy rate threshold to be 12\% and 64\% respectively to display the associated color information. Moreover we tested how accurately of our scheme can represent the true occupancy. Another 5-day field test was carried out from Monday to Friday after we set the scheme. We count it as an error if the true occupancy of the parking lot does not lie in the displayed range as expected. Thus the Mean Square Error (MSE) is presented here by comparing the true occupancy and color scheme info every 5 minutes from 8:30 am to 10:30 am. The resulting MSE is 11.67\%. Therefore, we are confident the algorithm is effective and can approximate the ground truth.
\subsection{ Searching Time}
\begin{figure}[!b!t!h]
\centering
\includegraphics[width=0.65\textwidth]{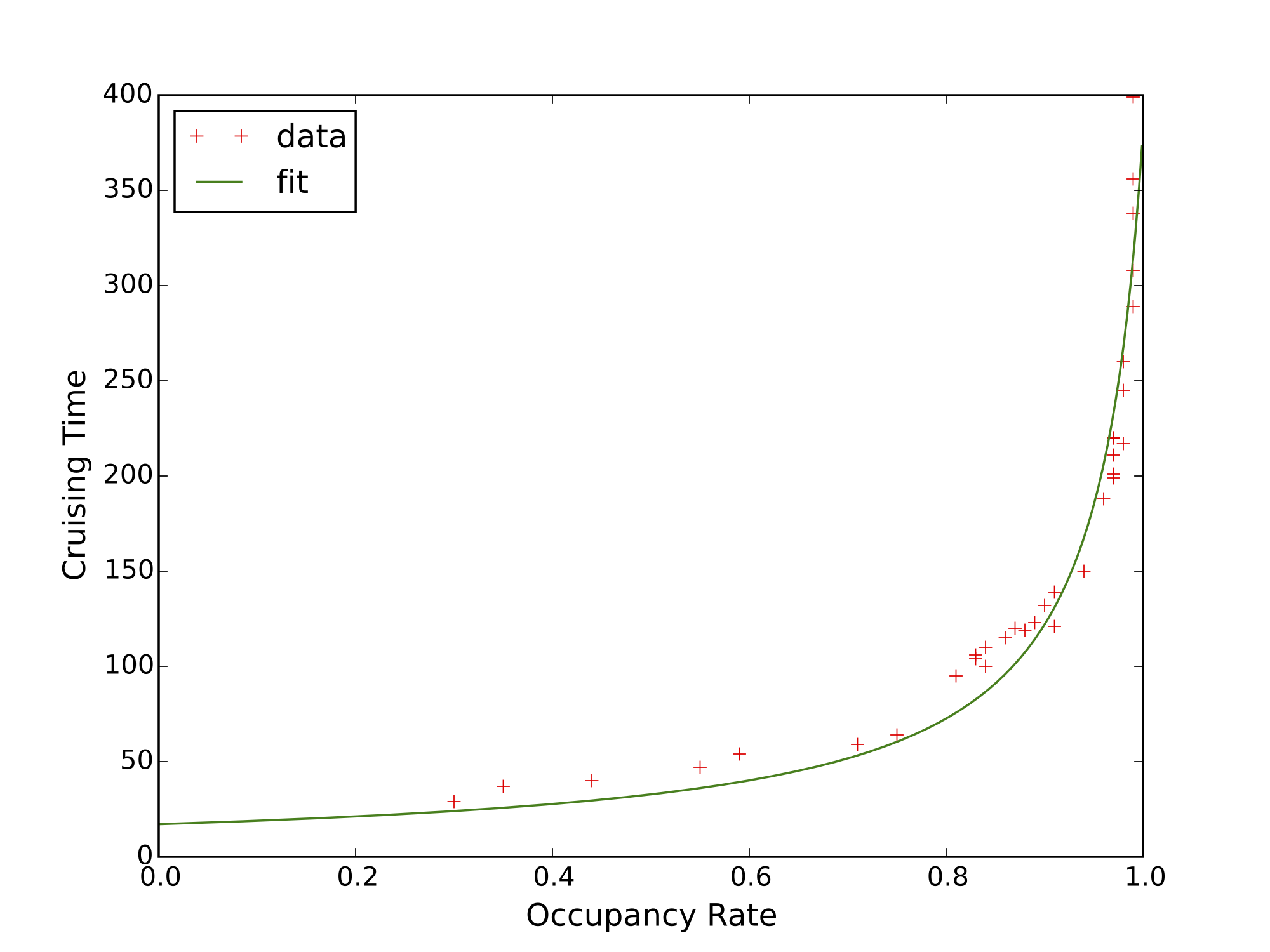}
\caption{Cruising time sample fitting curve, x-axis is the occupancy rate scaled from 0 to 1, y-axis is the cruising time in the unit of second}
\label{fig:ps1_searchTime_fitting}	
\end{figure}

Searching for a parking spot, especially during the peak hours, takes a significant portion of the time in the whole driving experience. Thus the searching time for parking, also known as cruising time, is another crucial piece of information for travelers. The cruising time is defined as a period from when a car enters the PS-1 to when the car is parked (if a parking spot is available) or the car leaves the PS-1 (if no parking spot is available). During the onsite survey, we recorded the cruising time with corresponding occupancy rate at the same time. Given the intuition that cruising time largely depends on the parking occupancy \cite{balijepalli2013calibration, belloche2015street}, we propose the following search time function,
\begin{equation}
	T = \frac{ \alpha }{1 - \beta r } \label{eq:ps1_searchTime_func}
\end{equation}
where $r$ is the occupancy rate, $T$ is the time of searching for a parking spot, and $\alpha$, $\beta$ are the function parameters. Equation (\ref{eq:ps1_searchTime_func}) is an extended form of the function proposed by Axhausen et al \cite{axhausen1994effectiveness} and the other relevant works \cite{panayappan2007vanet,kavaler2012micro,qian2014optimal,qian2015optimal}. The parameters $\alpha$ and $\beta$ are obtained by minimizing the sample fitting error ($l_2$ norm) as follows
\begin{align}
	\min & \sum_i^n \| \frac{ \alpha }{1 - \beta r^{(i)} } - T^{(i)} \|^2	\\
	s.t. {\quad} & \alpha \geq 0 \\ 	
	    & \beta \geq 0
\end{align}
where $r^{(i)}, T^{(i)}$ are the sample data from our survey. We use 5-fold cross validation to pick the value of the predictor’s parameters. Thus 80\% of the data is used for training and 20\% of the data remains for testing. As a result, Figure \ref{fig:ps1_searchTime_fitting} shows the fitted search time function with $\alpha$ = 17.2678, $\beta$ = 0.9946. The Mean Absolute Percentage Error (MAPE) for training and testing are 13.49\% and 15.29\% respectively. Given the known parameters $\alpha$ and $\beta$, we retain the search time when occupancy rate equals $r_1$ = 85\% and $r_2$ = 95\% using function (\ref{eq:ps1_searchTime_func}). The associated results of searching time are 108.89 seconds and 290.08 seconds respectively. For the simplicity of displaying the information, we round the time into minutes, which are 2 minutes and 5 minutes in corresponding scenarios.

\section{CONCLUSIONS}
This paper answers the following three questions. First, how can we sense parking occupancy in an efficient, accurate and cheap way in a multi-layer commuting parking structure? We solve this question by installing spot-by-spot parking sensors and accurately detecting the occupancy on the top floor. Second, how should the parking information be displayed for users to better understand and make parking choices? We answer this question by providing the web interfaces containing occupancy and search time information, along with a color-coding interface design. Third, what is the system architecture of a parking sensing and information system, and what is the best method to deploy it in the real world? We use Wireless Sensor Network technology that is battery-driven, power-saving, real-time and scalable. \\~\\
The sensing system was deployed at a parking garage on Stanford campus. It consists of 107 spot-by-spot micro-radar parking sensors, two access points and seven repeaters. It is integrated with a cloud-based data acquisition and visualization platform that is highly scalable de- pending on the web traffic. The communication between the WSN and cloud servers relies on a 4G/3G (forth- or third-generation) network. An informative web interface is also provided, which enables users to get both occupancy and parking search time information. Furthermore, a color scheme is designed in the system that allows users to have an intuitive perception of parking availability.\\~\\
Inferring the whole parking structure occupancy based on top floor measurements is reasonable when a parking structure is typically filled from the bottom to the top. Our assumption is made based upon the morning commute parking patterns. In the future, we plan to investigate the evening peak parking pattern as well, since all of the color-coding mechanisms and occupancy visualizations are configured according to the morning traffic pattern. It would have different threshold values for both the color scheme and occupancy rate in the evening. Another challenge is that, although the discussed parking occupancy pattern exists in most parking structures serving commuters, other types of parking flow patterns may be found in the structures serving tourists or visitors. In that case, we can expand our parking system by incorporating counting sensors at the entrances and exits of these parking structures. Furthermore, thanks to the scalable cloud configuration and open web interface, we plan to coordinate with Stanford’s parking enforcement agency by providing more customized services.

\bibliographystyle{unsrt}
\bibliography{ps1_axriv_main.bib}

\end{document}